\documentclass[aps,prl,twocolumn,showpacs]{revtex4}
\usepackage{amsbsy,latexsym}
\usepackage{amsfonts}
\usepackage{amssymb}
\usepackage[mathscr]{eucal}

\newcommand{\al}{\alpha}

\newcommand{\la}{\lambda}

\newcommand{\e}[1]{\exp\left(#1 \right)}

\newcommand{\ket}[1]{\vert #1 \rangle}
\newcommand{\bra}[1]{\langle #1 \vert}

\newcommand{\NN}{{\mathscr N}}

\begin{document}
\voffset 1.5cm
\title{Optimal scheme for estimating a pure qubit state via local measurements}

\author{E.~Bagan, M.~Baig, and R.~Mu{\~n}oz-Tapia}
\affiliation{Grup de F{\'\i}sica Te{\`o}rica \& IFAE, Facultat de
Ci{\`e}ncies, Edifici Cn, Universitat Aut{\`o}noma de Barcelona, 08193
Bellaterra (Barcelona) Spain}
%\date{06/05/2002}

\begin{abstract}
We present the optimal scheme for estimating a pure qubit state
by means of local measurements on  $N$ identical copies. We give
explicit examples for low $N$. For large $N$, we show that  the
fidelity saturates the collective measurement bound up to order
$1/N$. When the signal state lays on a meridian of the Bloch
sphere, we show that this can be achieved without classical
communication.
\end{abstract}

\maketitle

State estimation is a central topic in Quantum Mechanics. Quantum
measurements give only partial information about the state of the
system under consideration. If an unlimited number of copies of a
given state were available, one could in principle determine
exactly this state provided an infinite number of measurements
were performed. In the real world we have access to a limited
number of copies and time for a finite number of measurements,
and the best we can achieve is an estimate of the state. It is,
thus, of great importance to design optimal strategies which
maximize the knowledge one can acquire about a general quantum
state.

Over the last few years a great deal of work has been devoted to
the optimal estimation of states and many important features have
been recognized \cite{peres1,mp, state, us-peres,gm,matsumoto},
one of the most interesting ones being that collective
measurements (CM) are more informative than individual
ones~\cite{peres1,mp}. The experimental implementation of such
measurements seems, though, quite involved. In practice,
individual von Neumann measurements are far easier to perform.
Moreover,  one can show that optimal individual measurements are
of this type in the situations considered in this
letter~\cite{prep}. The most general individual measurement
procedures come under the name of LOCC (local operations and
classical communication) schemes. In this framework, one allows a
wide class of local operations for which, depending on the
outcome of the local measurement performed on a copy, appropriate
transformations can be applied on the subsequent copies of the
state before measuring again.
We consider the general situation in which the copies of the system may have never interacted in the past and may be macroscopically separated. Thus, classical communication becomes essential in these  schemes. This is in contradistinction to collective measurements, for which the role of classical communication is played by the quantum entanglement  of the measuring devices.
 In this LOCC context, some numerical
simulations and experimental tests have been performed recently
\cite{fisher,hannemann}.

In this letter we will deal with pure states of a single qubit
$\ket{\psi}$. To each one of these states it corresponds a unique
unit vector $\vec n$ on the Bloch sphere; i.e.,
$\ket{\psi}\bra{\psi}=(1+\vec{n}\cdot \vec{\sigma})/2$, where
$\vec{\sigma}$ are the usual Pauli matrices. We will focus on two
situations to which we will simply refer as 3D and 2D. In the
former, no prior knowledge of $\vec n$ is assumed, whereas for
the later (2D) $\vec n$ is known to lay on a meridian of the Bloch
sphere. Our aim is to design LOCC measurements such that we can
obtain the best estimate of $\vec n$ and, hence, of $\ket{\psi}$.
We will investigate how good these LOCC measurements are as
compared to the collective ones. For the latter, the mean
fidelity $F$ is commonly used as a figure of merit and many
results can be found in the literature ---in particular, the
large $N$ asymptotic behavior of $F$ is now known for the most
interesting approaches \cite{mp,us-peres}.

{}For $N$ identical copies of $\ket{\psi}$ optimal CM lead to a
fidelity that behaves as $F_{\rm CM}\approx 1-1/N$ for large $N$
in 3D~\cite{mp}. From~\cite{holevo,bu} one can likewise compute
the asymptotic fidelity in 2D, which is $F_{\rm CM}\approx
1-1/(4N)$. These results are the absolute upper bound for {\it
any} measurement scheme, therefore within the LOCC framework the fidelity
$F$  can not approach unity at a rate larger than $1-1/(4N)$ in 2D
($1-1/N$ in 3D)~\cite{gm}. In this letter we demonstrate that,
rather unexpectedly, this asymptotic behavior can be achieved in
2D with just local measurements and {\em no classical
communication}. In 3D classical communication seems necessary to
saturate the asymptotic CM bound, however, we have verified that
for the optimal approach the fidelity reaches the CM regime very
fast (for $N\gtrapprox 12$). Therefore, CM do not provide a
significant improvement over local measurements.

The estimation procedure goes as follows.  After the $N$
measurements (one on each copy) have been performed, a list of
outcomes is obtained, which we symbolically denote by $x$. Based
on $x$,  an estimate for $\ket{\psi}$ can be guessed,
$\ket{\psi_{\rm guess}(x)}$. The fidelity is then defined as the
overlap
\begin{equation}\label{f}
 |\bra{\psi}\psi_{\rm guess}(x)\rangle|^2={1+\vec n\cdot\vec M(x) \over
  2}\equiv f_n(x),
\end{equation}
where, as above,  $\vec{n}$ is the unit vector on the Bloch sphere
corresponding to the state $\ket{\psi}$ and $\vec{M}(x)$ is that
corresponding to $\ket{\psi_{\rm guess}(x)}$.  The average
fidelity can  be written as $
  F\equiv\langle f \rangle =
\sum_{x} \int dn\, f_n(x)\;P_{n}(x)$, where $P_{n}(x)$ is the
probability of getting the outcome $x$ if the Bloch vector is
$\vec{n}$, and  $dn$ is the measure on the sphere in 3D (on the
unit circle in 2D). Note that in both situations we have  assumed
no further prior knowledge of $\ket{\psi}$.

Any local  von Neumann measurement is represented by two
projectors $O(\pm \vec{m})=(1\pm \vec{m}\cdot \vec{\sigma})/2$,
where $\vec{m}$ is a unit Bloch vector  characterizing the
measurement (in a spin system, e.g., $\vec{m}$ is the orientation
of a Stern-Gerlach). We further note that for any unitary
transformation $U$ and for any $\vec m$, there is always a vector
$\vec m'$ such that $O(\vec m')=U O(\vec m) U^\dagger$. So, any
local operation on an individual copy of $\ket{\psi}$ may be
viewed as a redefinition of the Bloch vector that characterizes
the measurement which will be performed on that copy. We
hereafter adopt this point of view. For von Neumann measurements,
the set of outcomes $x$ can be expressed as a $N$-digit binary
number $x=i_{N}i_{N-1}\cdots i_{2}i_{1}$, were $i_k$ ($=0,1$)
indicates that upon  measuring on the $k$'th copy, this is
projected on the $O[(-)^{i_k}\vec{m}_k]$ projection space. The
most general local measurement is realized when we allow
$\vec{m}_{k+1}$ to depend also on the list of  previous outcomes
$i_{k} i_{k-1}\cdots i_{2}i_{1}\equiv x_{k}$ (hence, $x=x_N$). We
thus write $\vec{m}(x_k)$ instead of $\vec{m}_k$.  Note that
$\vec m(x_{k})$ satisfy
\begin{equation}\label{von-neumann}
  \vec m(1x_{k-1})=-\vec m(0x_{k-1}).
\end{equation}
The fidelity of a general LOCC scheme can be written as
\begin{equation}
    F=\sum_{x=00\cdots0}^{2^{N}-1}\int dn{1+\vec n\cdot\vec M(x)\over2}
    \prod_{k=1}^{N}{1+\vec n\cdot\vec m(x_{k})\over2},
    \label{fidelity-2}
\end{equation}
where the last product is  the probability $P_n(x)$.

The optimal LOCC scheme is the one that
maximizes~(\ref{fidelity-2}) over the Bloch vectors
$\vec{m}(x_k)$ and $\vec{M}(x)$. Using the Schwartz inequality,
it is straightforward to see that the best guess $\vec M(x)$ must
be proportional to the vector
\begin{equation}
   \vec V(x)=\int dn\,\vec n \prod_{k=1}^{N}{1+\vec n\cdot\vec
   m(x_{k})\over2},
    \label{vector-1}
\end{equation}
i.e., $\vec{M}(x)=\vec{V}(x)/|\vec{V}(x)|$. In this case, the
maximum fidelity reads
\begin{equation}
    F={1\over2}\left(1+\sum_{x}|\vec V(x)|\right).
    \label{fidelity-general}
\end{equation}
For a fixed set of measurements $\{m(x_k)\}$ (optimal or not) and a
given set of outcomes $x$, the
guess (\ref{vector-1}) provides the best estimate of the signal
state. This simple and general result does not seem to be
conveyed in the literature. We next show how~(\ref{vector-1}) can
be used to improve the quality of state estimation schemes based
on local measurements.

 Consider $N=2\NN$ copies of the state $\ket{\psi}$ whose vector
 $\vec n$ is known to be on the equator ($xy$-plane) of the Bloch
 sphere (2D case).
 Since the expectation value of $\vec{\sigma}$ is
$\bra{\psi}\vec{\sigma}\ket{\psi}=\vec{n}$, the central limit
theorem strongly suggests to adopt the following scheme. Let
$\vec{e}_1$, $\vec{e}_2$ be the two unit vectors pointing along
the $x$ and $y$ axes respectively. For each $i$ ($=1,2$) perform
$\NN$ measurements of  the observable $\sigma_{i}=
O(\vec{e}_i)-O(-\vec{e}_i)$. Assume we have obtained $\NN
\alpha_i$ times the outcome $+1$ (consequently,  $\NN
(1-\alpha_i)$ times the outcome $-1$). One is driven to propose $
\vec M\propto\sum_{i}\alpha_i \vec e_{i}+ \sum_{i}
(1-\alpha_i)(-\vec e_{i})$ (the mean value of these outcomes) as
the Bloch vector of $\ket{\psi}$. More precisely,
\begin{equation}\label{cl-guess}
  M_i(\alpha)=\frac{
  2\alpha_i-1}{\sqrt{\sum_j (2\alpha_j-1)^2}}.
\end{equation}
By doing so, the limiting behavior $\vec M\stackrel{\NN\to
\infty}{\longrightarrow}\bra{\psi}\vec{\sigma}\ket{\psi}=\vec n$
is ensured by the central limit theorem. With the techniques
described below one can show that the fidelity for the
guess~(\ref{cl-guess}) has the asymptotic expression
\begin{equation}
    F_{\rm CL}=1-{3\over 8}{1\over N}+\dots,
    \label{fidelity-cl}
\end{equation}
where ${\rm CL}$ stands for central limit.

According to~(\ref{vector-1}), however, this guess cannot be
optimal. Let us show that indeed this scheme can be improved
using~(\ref{vector-1}) instead of~(\ref{cl-guess}). In this
particular situation Eqs.~\ref{fidelity-general}
and~\ref{vector-1} read $F_{\rm OG} =(
1+\sum_{\alpha}|\vec{V}(\alpha)|)/2$ (OG stands for optimal guess)
and
\begin{eqnarray}\label{m-guess}
  \vec{V}(\alpha)&=&\prod_{i=1}^2\pmatrix{\NN\cr \NN\al_{i}}
  \int dn\; \vec n \nonumber \\
    & \times &\prod_{j=1}^2\left({1+n_{j}\over2}\right)^{\NN\al_{j}}
     \left({1-n_{j}\over2}\right)^{\NN(1-\al_{j})},
\end{eqnarray}
where  we have used the shorthand notation
$\alpha=(\alpha_1,\alpha_2)$. To obtain the large $N$ limit of
$F_{\rm OG}$ we use the formula
\begin{eqnarray}
&&  \pmatrix{\NN\cr\al \NN} q^{\al \NN} (1-q)^{(1-\al)\NN}\to
    {1\over\sqrt{2\pi \NN q(1-q)}}\nonumber\\
&&\quad \times\e{-{\NN\over 2}{(\al-q)^2\over q(1-q)}}
    \left\{1+O(1/\sqrt \NN)\right\} ,
    \label{stirling}
\end{eqnarray}
with $q=(1+n_{j})/2$, and approximate the sum over the outcomes
$\alpha$ by an integral. Performing the change of variables
$r_i=2\alpha_i-1$, we get
\begin{equation}\label{fidelity-fixed-2}
  \sum_{\al}|\vec{V}(\alpha)| \simeq
    \int d^2r\left|\vec{V}\right|
     \equiv  \int d^2r\left|\int dn\, \vec{Q}{\rm e}^{-{\NN\over 2}E}\right|,
\end{equation}
with $E=\sum_{i=1}^2{(n_{i}-r_{i})^2/(1-n_{i}^2)}$ and $\vec{Q}=P
\,
  \vec{n}=({\NN/2\pi}){\prod_{i=1}^2(1-n_{i}^2)^{-1/2}}\;\vec{n}$.
As it will become clear below, the terms of order $1/\sqrt \NN$
neglected in these approximations give no contribution to the
final answer. The integral (\ref{fidelity-fixed-2}) will now be
evaluated with the aid of the saddle point technique. We use
spherical coordinates and write $\vec n=(\cos\phi,\sin\phi)$,
$\vec r=r\;(\cos\gamma,\sin\gamma)$. For large $\NN$, the
integrand peaks at $r=1$, which suggests to  write $r=1+\xi$. We
denote by $\phi_{m}$ or, equivalently, by $\vec n_{m}$, the
location of the minimum of $E$, which can be computed as a power
series in $\xi$ of the form
$\phi_{m}=\gamma+\la_{1}\xi+\la_{2}\xi^2+\dots$. We next expand
$E$ and $\vec{Q}$ as a power series in $t\equiv\phi-\phi_{m}$ and
write $\vec{Q}=\vec{L}+\vec{M} t +\vec{R} t^2+\dots$. The
exponential in~(\ref{fidelity-fixed-2}) is seen to have the form
\begin{equation}\label{exponent'}
{\rm e}^{-\frac{\NN}{2}E}={\rm e}^{-{\NN\over 2}A}{\rm
e}^{-{\NN\over 2}B t^2 }{\rm e}^{-{\NN\over
    2}\times O(t^3)},
\end{equation}
where $2 B$ is the second derivative of $E$ with respect to $t$,
and the last exponential factor  can also be expanded in powers
of $t$. Neglecting contributions which vanish exponentially in
$\NN$, one has
\begin{equation}
\vec{V}=    \int dn\, \vec{Q} {\rm e}^{-{\NN\over 2}E} = {{\rm
e}^{-{\NN\over 2}A}\over \sqrt{2\pi \NN B} } \left(\vec{L}+ {
\vec{R}\over
     \NN B}+\dots
\right),
    \label{ebc-6}
\end{equation}
where $dn=d\phi/(2\pi)=dt/(2\pi)$. At this point all the
coefficients can be expanded in powers of $\xi$, e.g.,
$B=B_{0}+B_{1}\xi+\dots$, $R=R_{0}+R_{1}\xi+\dots$, and so on.
Note however that the expansion of $A$ starts at order $\xi^2$,
i.e., $A=A_{2}\xi^2+A_{3}\xi^3+\dots$. For $\vec{Q}=\vec n P=\vec
n_{m} P+(\vec n-\vec n_{m})P$, we readily note that~(\ref{ebc-6})
yields
\begin{equation}
\vec V=\vec n_{m} \int dn\,P{\rm e}^{-{\NN\over2}E}+{\vec I\over
\NN},
    \label{ebc-7}
\end{equation}
where $\vec I/\NN$ is given by~(\ref{ebc-6}) with $\vec{Q}=(\vec
n-\vec n_{m})P$. We have written a factor $1/\NN$ in the last term
to make explicit that this contribution is of order $\NN^{-1}$.
Hence
\begin{equation}
    \left|\vec V\right|=\int dn\,P{\rm e}^{-{\NN\over2}E}+{\vec n_{m}
    \cdot\vec I\over \NN} + O\left({1\over \NN^2}\right)
    \label{ebc-8}
\end{equation}
and we finally have up to $O(1/\NN^2)$
\begin{equation}
    \int d^2r|\vec V|=1-\!\int\! d^2r \! \int \!dn
    P(1-\vec n\cdot\vec n_{m}){\rm e}^{-{\NN\over2}E}.
    \label{ebc-9}
\end{equation}
The leading term (unity) comes from the first integral in
(\ref{ebc-8}), which is straightforward to compute exchanging the
order of integration. Note also that the integral of the leading
term on the right hand side of~(\ref{stirling}), $\int d^2 r P
\exp(-\NN E/2)$, is one, and so is the sum over $\alpha$ on the
left hand side \emph{independently of $\NN$}. This shows that the
terms of order $1/\sqrt \NN$ neglected in~(\ref{fidelity-fixed-2})
indeed cancel. The subleading term in~(\ref{ebc-9}) can  be
computed using the general expression~(\ref{ebc-6}) where now
$Q=(1-\vec n\cdot\vec n_{m})P=R t^2+O(t^3)$ (i.e., $L=M=0$).
Moreover, only the leading terms in powers of $\xi$ have to be
retained, namely, $A_2$, $B_0$ and $R_0$ (in particular, we  just
need $\lambda_1$). This is so because effectively
$O(t^2)=O(\xi^2)=O(1/\NN)$. The integral in~(\ref{ebc-9}) is then
\begin{eqnarray}
&&\int_{0}^{2\pi} {d\gamma\, R_0 \over\sqrt{2\pi \NN^3 B_{0}^3}}
    \int_{0}^{\infty} dr\, r\exp\left\{-{\NN A_{2}\over
2}(r-1)^2\right\}\nonumber\\
&&= \frac{2 \pi}{\NN}\int_{0}^{2\pi}{d\gamma\over 2\pi}{R_0 \over
\NN\sqrt{A_{2} B_{0}^3}},
    \label{ebc-10}
\end{eqnarray}
where again terms vanishing exponentially as $\NN\to\infty$ have
been neglected. The coefficient $\lambda_1$ is computed to be
$\la_{1} =-\cot2\gamma$. With this, $B_{0}=2$,
$R_{0}=(\NN/2\pi)\csc 2\gamma$ and $A_{2}=2\csc^2 2\gamma$. The
subleading term in (\ref{ebc-9}) yields $-1/(4\NN)$, and
substituting back in~(\ref{fidelity-general}) we finally
obtain
\begin{equation}
    F_{\rm OG}=1-{1\over4}{1\over N}+\dots
    \label{fidelity-modul}
\end{equation}
(recall that $N=2\NN$). As announced, this fidelity is larger than
$F_{\rm CL}$. Moreover, it saturates the absolute upper bound
given by CM even though classical communication has not been
used. In this sense, the behavior of the most basic local scheme
(without classical communication) in 2D is qualitatively similar
to that obtained from CM provided the optimal
guess~(\ref{vector-1}) is used. Working along the same line, we
have computed the asymptotic fidelities $F_{\rm CL}$ and $F_{\rm
OG}$ in 3D. One has
\begin{equation}
    F_{\rm CL}=1-{6\over5}{1\over N}+\cdots;\quad
    F_{\rm OG}=1-{13\over12}{1\over N}+\cdots,
    \label{F's in 3D}
\end{equation}
where  again  we note that $F_{\rm OG}>F_{\rm CL}$. Despite this
improvement, the CM bound $F_{\rm CM}=1-1/N$  is not saturated,
although the subleading term of $F_{\rm OG}$ is only 8\% less
than that of $F_{\rm CM}$.

So far classical
communication has not been exploited,  i.e., the Bloch vectors
$\vec{m}(x_k)$ were both $x$-independent and non-optimal. We now turn
to the full-fleshed LOCC schemes. Hereafter only the general case 3D
will be considered.

The first non trivial case is $N=2$. Here, $x$ takes four
possible values, $00$, $01$, $10$ and $11$. There are three
independent vectors, namely, $\vec{m}(0),\vec{m}(00),\vec{m}(01)$
(the other three are obtained using Eq.~\ref{von-neumann}). The
first vector, $\vec{m}(0)$, is  arbitrary and we take
$\vec{m}(0)=\vec{z}$. The optimal fidelity is obtained by
maximizing (\ref{fidelity-general}) with respect to $\vec{m}(00)$
and $\vec{m}(01)$. A straightforward calculation gives
\begin{equation}\label{N=2}
\sum_{x}|\vec V(x)|=\frac{1}{6}\sum_{k=0,1}\left(
\big|\sin\frac{\theta_{0k}}{2}\big|+\big|\cos\frac{\theta_{0k}}{2}\big|
\right),
\end{equation}
where $\theta_x$ is the polar angle of the vector $\vec{m}(x)$.
The maximal value of (\ref{N=2}) is attained for
$\theta_{00}=\theta_{01}=\pi/2$. Notice that $\theta_{00}$ and
$\theta_{01}$ are maximized independently, so $\vec{m}(00)$,
$\vec{m}(01)$ do not need to be equal; they are only required to
be orthogonal to $\vec{m}(0)$. Substituting back
in~(\ref{fidelity-general}) we find $F^{(2)}=(3+\sqrt{2})/6$ (see
also~\cite{fisher}).
{}From  Eq.~\ref{vector-1}
we obtain the optimal guess
\begin{equation}\label{guess-N=2}
 \vec{
 M}^{(2)}(x)=\frac{\vec{m}(x_2)+\vec{m}(x_1)}{\sqrt{2}}=\vec{s}(x),
\end{equation}
hence, e.g., $\vec{M}^{(2)}(01)=[\vec{m}(01)+\vec{m}(1)]/\sqrt{2}
=[\vec{m}(01)-\vec{m}(0)]/\sqrt{2}$.

The case $N=3$ is very similar. The optimal Bloch vectors,
$\vec{m}(x_1)$, $\vec{m}(x_2)$, $\vec{m}(x_3)$, are found to
be mutually orthogonal.
Since there is no further constrain, one can choose the three fixed
(i.e., independent of $x$)
vectors $\vec m(x_{k})=\vec e_{k}$, $k=1,2,3$. This shows that for
$N=3$ (as well as for $N=2$)
the optimal estimation schemes based on local measurements do not
require classical communication.
For each outcome $x$ the optimal guess is
$\vec M^{(3)}(x)=[\vec{m}(x_3)+\vec{m}(x_2)+\vec{m}(x_1)]/\sqrt{3}$,
which is a straightforward generalization of $\vec M^{(2)}(x)$,
and
yields $F^{(3)}=(3+\sqrt{3})/6$. These  results
could somehow be anticipated:
if $O(\vec m)\ket{\psi}\not=0$ we can only be sure
that the Bloch
vector of $\ket{\psi}$ is not $-\vec m$. Intuition suggest to use
the subsequent copies of $\ket{\psi}$ to explore the plane orthogonal to
$\vec m$. Thus, the optimal Bloch vectors $\vec m(x_{k})$ tend to be mutually
orthogonal.

The case $N=4$ is more complex, since four mutually orthogonal
vectors cannot fit onto the Bloch sphere.
%We now have to optimise 14
% vectors,
% which can be gruped in two independent families of seven.
% With such a large number of vectors an analytical calculation is too
% involved and
% we have resorted to a numerical optimization.
The solution exhibits some interesting features. First, the
optimal Bloch vectors now depend on the outcomes of the previous
measurements. Therefore classical communication does play a crucial
role for $N>3$.  However, $\vec{m}(x_1)\perp \vec{m}(x_2)$ and, as
before, one can choose $\vec{m}(x_i)=\vec e_{i}$, for $i=1,2$. Only
for the third and fourth measurement one really has to take different
choices
in accordance to the sequence of the preceding outcomes.
The Bloch vectors of the third measurement can be parametrized by a
single angle $\alpha$ as
$\vec{m}(x_3)=\cos\alpha \, \vec{u}_1(x)+\sin\alpha \, \vec{v}_1(x)$,
where $\vec{u}_1(x)=\vec{m}(x_1)\times \vec{m}(x_2)$ and
$\vec{v}_1(x)=\vec{u}_1(x)\times \vec{s}(x)$ and $\vec{s}(x)$ is defined
in~(\ref{guess-N=2}). The optimal value of this angle is
$\alpha=0.502$. We cannot give any insight as to why this value is
optimal. However, in agreement with our
intuition, $\vec
m(x_{3})\perp \vec s(x)$, i.e., the third measurement probes the
plane orthogonal to the Bloch vector one would guess from the first
two outcomes.
Two angles are required to parametrize the
vectors of the fourth measurement. They are given by
$\vec{m}(x_4)=\cos\gamma \, \vec{u}_2(x)+\sin\gamma \, \vec{v}_2(x)$,
where $\vec{u}_2(x)= \vec{s}(x)\times \vec{m}(x_3)$,
   $\vec{v}_2(x)= \cos\beta \, \vec{m}(x_3)-\sin\beta\,
  \vec{s}(x)$.
The optimal values of these angles are  $\beta=0.584$,
$\gamma=0.538$, %(see Fig.~1),
and the corresponding fidelity is $F^{(4)}=0.8206$. This is just
$1.5\%$ lower than the absolute bound $F_{\rm
CM}^{(4)}=5/6=0.8333$ attained with CM. We also give the values
of the maximal LOCC fidelities for $N=5,6$. They are
$F^{(5)}=0.8450$ and $F^{(6)}=0.8637$. It is interesting to note
that for $N>3$, it pays to relax optimality at each step. Hence,
one-step adaptive schemes~\cite{fisher,hannemann} are not
optimal, though the differences are very small; e.g., for $N=4$,
$F^{(4)}>F^{(4)}_{\mbox{\scriptsize
adaptive}}=(15+\sqrt{91})/30\approx 0.8180$.

Having learnt from the low $N$ cases, we are in the right position
to compute the asymptotic fidelity of this scheme. For that, we
take inspiration in variational methods as follows. Suppose we
have performed a large number $N_{0}=\sqrt{N}$ of measurements and
obtained the guess $\vec M_{0}$. It is clear that the subsequent
$2\bar N=N-N_{0}$ guesses will hardly differ from $\vec M_{0}$.
We hence substitute in~(\ref{fidelity-2}) the ansatz $\vec
M(x)\approx \vec M_{0}\cos\omega+\sin\omega(\vec u \cos\tau+\vec
v \sin\tau)$, where $\vec u$, $\vec v$ are two unit vectors which
along with $\vec M_{0}$ form an orthogonal bases, $\omega =\la
\sqrt{ (2 \alpha_{u}-1)^2+(2 \alpha_{v}-1)^2}$, $\tan\tau=(2
\alpha_{v}-1)/(2 \alpha_{u}-1)$, and $\lambda$ is a variational
parameter. As above, $\bar N\alpha_{u}$ ($\bar N\alpha_{v}$) is
the number of times we obtain the outcome $+1$ when we measure
$\vec\sigma\cdot\vec u$\quad ($\vec\sigma\cdot\vec v$). Note that
in average $\omega$ will be small since we expect
$\alpha_{u,v}\approx 1/2$, and we need to retain terms up to order
$\omega^2$. Putting all this together one gets
from~(\ref{fidelity-2})
\begin{equation}
    F\gtrapprox1-(1-\la)^2(1-F_{0})-{\la^2\over N-N_{0}}+\cdots,
    \label{F-F0}
\end{equation}
where $F_{0}$ is the optimal fidelity for $N_{0}$ measurements and
the dots stand for subleading terms in inverse powers of $N$ and
$N_{0}$. We readily see that the optimal choice is $\la=1$, which
leads to $F\approx1-1/N$~\cite{gm2}. Hence, our LOCC scheme does
saturate the CM bound. Furthermore, numerical analysis reveals
that the CM regime is reached for values of $N$ as low as $12$.

In summary, we have obtained the optimal LOCC estimating scheme
for general qubit (pure) states and shown that its fidelity
saturates the collective measurement bound. For states that are
known to lay on a meridian of the Bloch sphere (2D case) we have
explicitly given a scheme whose fidelity saturates this bound
without invoking classical communication.

We acknowledge financial support from CIRIT project SGR-00185, Spanish Ministry of Science and Technology project BFM2002-02588,  and the European Funds for Regional Development (FEDER).
.


\begin{thebibliography}{99}
\bibitem{gm2} An alternative proof of this asymptotic limit
              based on the quantum Cramer-Rao bound and one step
              adaptive scheme can be found in \cite{gm} (see also
              \cite{matsumoto}).
\end{thebibliography}
\end{document}